\documentclass[a4paper,11pt]{article}
\usepackage{pos}
\usepackage{graphicx}
\usepackage{wrapfig}

\title{Neutrino Education, Outreach, and Communications Activities: Captivating Examples from IceCube}
 \ShortTitle{IceCube E\&O}

\author{The IceCube Collaboration \\{\normalsize \normalfont(a complete list of authors can be found at the end of the proceedings)}}

\emailAdd{madeleine.okeefe@icecube.wisc.edu}
\emailAdd{joc.argueta@gmail.com}
\emailAdd{ellen.bechtol@icecube.wisc.edu}
\emailAdd{jim.madsen@icecube.wisc.edu}
\emailAdd{katherineshirey@gmail.com}

\abstract{The IceCube Neutrino Observatory at the South Pole has tremendous emotional appeal---the extreme Antarctic environment coupled with the aura of a pioneering experiment that explores the universe in a new way. However, as with most cutting-edge experiments, it is still challenging to translate the exotic, demanding science into accessible language. We present three examples of recent successful education, outreach, and communication activities that demonstrate how we leverage efforts and sustain connections to produce engaging results. First we describe our participation in the PolarTREC program, which pairs researchers with educators to provide deployments in the Antarctic, and how we have sustained relationships with these educators to produce high-quality experiences to reach target audiences even during a pandemic. We then focus on three activities from the past year: a summer enrichment program for high school students that was also modified for a 10-week IceCube after school program, a virtual visit to the South Pole for the ScienceWriters 2020 conference, and a series of short videos in English and Spanish suitable for all ages that explain traveling, living, and working at the South Pole.

\vspace{4mm}
{\bfseries Corresponding authors:}
Madeleine O'Keefe$^{1*}$, Jocelyn Argueta$^{2}$, Ellen Bechtol$^{1}$, Jim Madsen$^{1}$, Katherine Shirey$^{3}$\\
{$^{1}$ \itshape University of Wisconsin--Madison}\\
{$^{2}$ \itshape PolarTREC}\\
{$^{3}$ \itshape eduKatey, LLC}\\[4mm]
$^*$ Presenter

\FullConference{37$^{\rm{th}}$ International Cosmic Ray Conference (ICRC 2021)\\
		July 12th -- 23rd, 2021\\
		Online -- Berlin, Germany}

}

\begin{document}
\maketitle

\section{Introduction}

The IceCube Neutrino Observatory is a multifaceted, cutting-edge astroparticle physics facility located at the South Pole. It has enormous appeal to lay audiences because of its size, science, and extreme location. It is, literally, one of the coolest experiments on Earth: the cubic-kilometer array of detectors is buried a mile under the ice at the South Pole, a cold, dry, desolate, and remote environment with only one sunrise and sunset per year.

The extreme polar environment that makes IceCube so fascinating also makes it inaccessible to the public. Visiting the South Pole is essentially impossible for our audiences, so the challenge of the IceCube communication, education \& outreach team is to find ways to bring the South Pole ``up north'' to convey the excitement, innovation, and adventure happening at the bottom of the world. For years, we have leveraged our partnership with the PolarTREC program, which pairs educators with researchers in the polar regions, to get IceCube science to audiences beyond the research community.

\section{PolarTREC}\label{sec:Polartrec}

\begin{wrapfigure}{r}{0.25\textwidth}
    \centering
    \includegraphics[width=0.23\textwidth]{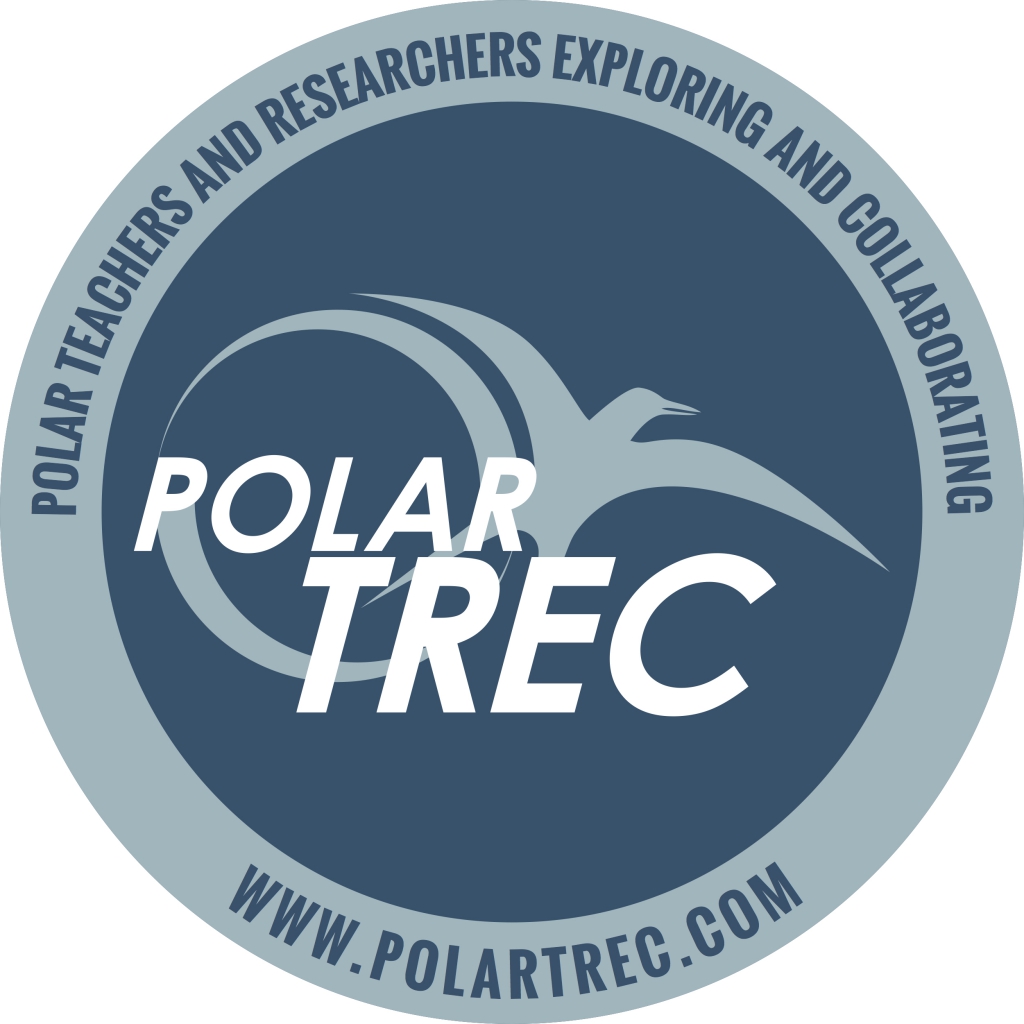}
    \label{fig:polartrec}
\end{wrapfigure}

PolarTREC and its predecessor TEA (Teachers Experiencing Antarctica and the Arctic) are US National Science Foundation programs that pair educators with polar researchers.  The educators become part of the research team, and their participation includes a field deployment.  IceCube and its predecessor AMANDA (Antarctic Muon and Neutrino Detector Array) have worked with nine educators since 2001, integrating them into activities at the South Pole for approximately three weeks with an additional week or so of travel there and back. At the South Pole, they get hands-on experience with experimental science and learn the wide range of tasks it takes to keep IceCube functioning. In turn, IceCube benefits from the fresh eyes of experienced educators who are able to convey the science and South Pole work in engaging, accessible ways.   

\section{IceCube for High School Students}

Since 2006, IceCube has provided a 2-week residential summer science enrichment program for students in grades 9-12 through the University of Wisconsin-River Falls (UWRF) Upward Bound program.  Upward Bound is a US federally funded program that provides support to help low-income students better prepare for success in pursuing an undergraduate degree after finishing high school. The students are from families with low incomes and/or households with parents that do not have a college degree.  The highly diverse group of students in the UWRF program attend Washington Technology Magnet High School in St. Paul, MN.

The number of students varies from year to year, ranging from 30 to 80. Smaller groups of students might represent only two grades, but for years with large numbers of participants, all four grades are represented in the student group. We rely on the PolarTREC/TEA educators to produce a meaningful, motivating educational experience.  Each summer, a new program has been developed that teaches science in a manner not possible in a traditional high school classroom.  Students learn the process of science by designing and creating projects, ranging from making technology-enhanced clothing to producing computer apps that convey the impact of the pandemic on their lives.  Along the way, they learn how to develop a preliminary design, produce a prototype, give and receive constructive feedback, and address challenges to make progress.  This approach was developed by longtime high school teachers and IceCube/AMANDA collaborators Eric Muhs and Steve Stevenoski, and more recently refined by Dr. Katherine Shirey.  

Dr. Shirey developed an educational activity using a computer-coding platform (https://code.org/) for the summer 2020 UWRF Upward Bound program \cite{UB2020} and subsequently leveraged this effort to devise a ten-week virtual after-school program for another group of high school students (interns) the following winter. 
IceCube After School was designed to layer three threads of learning: IceCube and neutrino science, programming, and computational thinking. Students wove these three threads into a final coded project to demonstrate their learning about IceCube science. To teach the interns about IceCube and related science, eight speakers provided guest lectures on their IceCube work. Topics ranged from neutrino oscillations to cosmic ray detection to science communication. The talks included a description of the presenters' careers so students could appreciate the multiple paths available to success as well as the variety of careers that involve physics. Several speakers emphasized how their career journeys shifted as they matured, and they encouraged the interns to imagine themselves as scientists even if they felt unsure about how to get there. 

To teach the interns about computational thinking and coding, short (15-20 minute) introductions were provided on topics (data types, loops, operators, and conditionals), and code.org activities were assigned for asynchronous completion to practice using concepts. Interspersed in the homework were engineering design stages requiring students to grapple with the challenge of creating a coded communication tool to teach a particular stakeholder  something about IceCube. Students identified peers, family, and teachers as stakeholders and crafted apps, animations, games, and interactive presentations in code.org to share their learning. Some students opted for alternatives to code.org, including compiled Python codes, storyboards, and websites. 

\begin{figure}[h!]
    \begin{center}
    \includegraphics[width=0.8\textwidth]{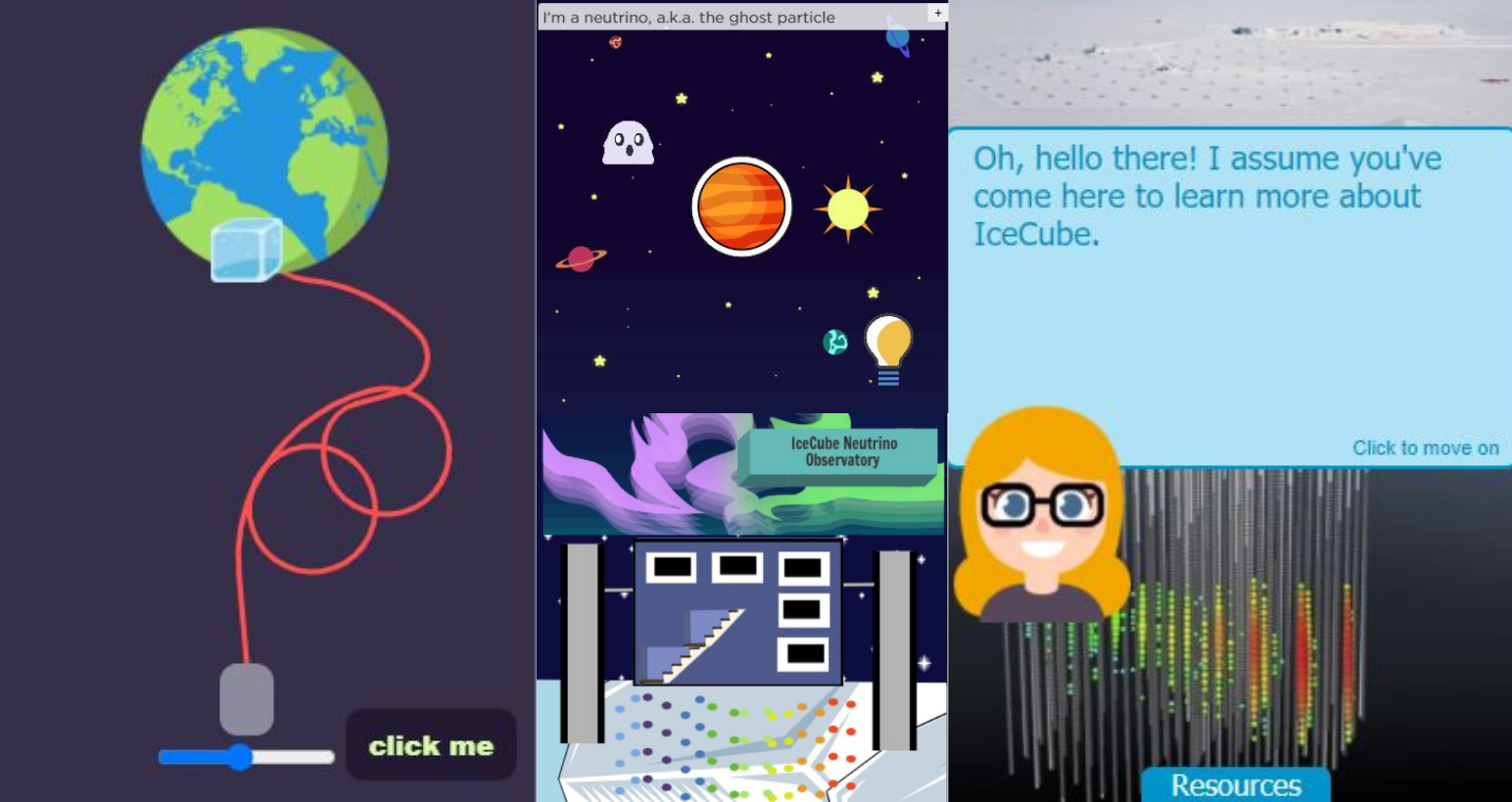}
    \caption{Examples of four students' final projects from the IceCube After School program.}
    \label{fig:IAS}
    \end{center}
\end{figure}

On the final day of the course, students presented their designs to an audience including parents, the guest speakers, and other IceCube collaborators in an online showcase. Each student made a slide to describe their work, posted a link to their IceCube animation code, and received comments from the virtual audience using Google Slides. \cite{IAS21}

\section{IceCube for Science Writers}

Like the vast majority of conferences and meetings in 2020, the annual ScienceWriters conference was organized virtually in response to the coronavirus pandemic. A joint meeting of the National Association of Science Writers (NASW) and the Council for the Advancement of Science Writing (CASW), ScienceWriters (SciWri) attracts hundreds of scientists, journalists, and science communicators every year. The conference comprises a mix of professional development workshops, briefings on the latest scientific research, extensive networking opportunities, and (usually) field trips---described as ``for science writers, by science writers.''

In lieu of the myriad in-person activities typically organized by the host institution of the conference, CASW called on attendees to create a virtual, Zoom-able experience about the science happening in their part of the world. It would ``provide a unique opportunity to travel virtually to the many institutions ScienceWriters may never be able to visit,'' said CASW in their call for proposals.

It was the perfect opportunity to showcase IceCube.

Madeleine O'Keefe, IceCube communications manager and NASW member, submitted a proposal to give a virtual tour of IceCube and introduce our South Pole experiment to SciWri20. When the proposal was accepted, we designed an hour-long virtual presentation with three main parts: an introduction to IceCube and our science, a prerecorded tour of the South Pole facilities, and a demo of a new augmented reality app ICEcuBEAR \cite{icecubear}  that takes users ``under the ice,'' designed by IceCube collaborator Dr. Lu Lu. Time was reserved at the end of the presentation for questions from attendees.

Finding someone to host the first segment and introduce IceCube was an important task. With diversity and inclusion being central to both IceCube's and CASW's missions, we knew we wanted someone who not only had the presence and skill to engage an audience of scientists and science writers in no more than 15 minutes but also offered a fresh voice to the overwhelmingly white and male-dominated field of physics. Fortunately, we had an ideal representative already part of the IceCube family: our 2019 PolarTREC educator, Jocelyn Argueta. \cite{JA}

\begin{wrapfigure}{l}{0.33\textwidth}
    \centering
    \includegraphics[width=0.33\textwidth]{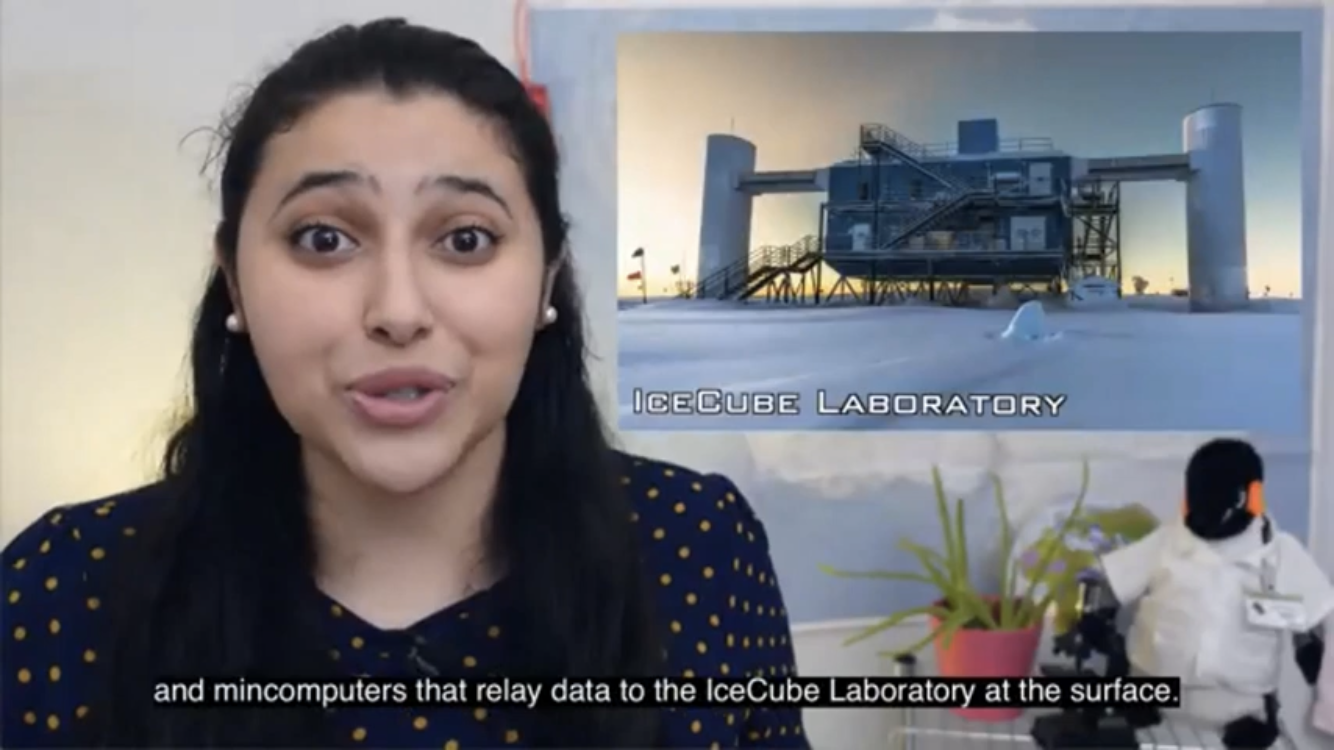}
    \caption{Screenshot from Jocelyn Argueta's video for IceCube's presentation at ScienceWriters 2020.}
    \label{fig:JA_SWYA}
\end{wrapfigure}

With her background in both science and performance, Argueta was already a seasoned science communicator. Based in La Mirada, California, Argueta has her own show where she stars as ``Jargie the Science Girl,’’ an energetic and charismatic scientist who teaches her audiences of schoolchildren about science through live experiments and more. In addition, Argueta had been helping with virtual science outreach throughout the pandemic, including a live public webcast to the South Pole \cite{JAwebcast} and her own series of videos, ``Tiny Ice’’ (see Section \ref{sec:TinyIce}). 

For the SciWri20 presentation, our whole team collaborated on a script that described our experiment and gave an accessible explanation of IceCube science.  Argueta used that script and various IceCube-provided graphics to produce the introductory video \cite{JAintro}, which we played during the presentation, ``Exploring the Cosmos from the South Pole Ice: A Virtual Tour of the IceCube Neutrino Observatory,'' on Thursday, October 8, 2020. It was supplemented by a prerecorded tour of the IceCube Laboratory at the South Pole by our winterovers, Dr. John Hardin and Dr. Yuya Makino \cite{WOtour}, and a live demo of the IceCube AR app by O'Keefe. Over 100 people attended and we received overwhelmingly positive feedback. 
The recorded presentation is available to watch on YouTube, where it has been viewed more than 7,500 times. \cite{swya}

\section{IceCube for All}\label{sec:TinyIce}


As described in Section \ref{sec:Polartrec}, the PolarTREC program matches educators with researchers in the polar regions and deploys them to the field to gain hands-on research experience for 4-6 weeks. Educators supported by PolarTREC share their experiences through blog posts and a live webinar. Upon their return, all educators generate a lesson plan or final project unique to the research they participated in as well as develop outreach events in their communities.

\begin{wrapfigure}{r}{0.33\textwidth}
    \centering
    \includegraphics[width=0.33\textwidth]{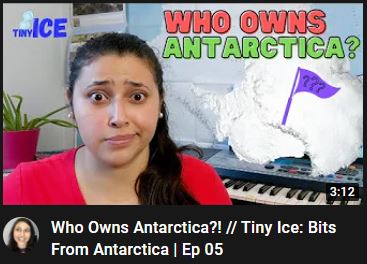}
    \caption{Thumbnail for episode 5 of ``Tiny Ice'' produced by Jocelyn Argueta.}
    \label{fig:TinyIce}
\end{wrapfigure}

In 2019, the IceCube Neutrino Observatory hosted PolarTREC educator Jocelyn Argueta, who returned from her expedition in early 2020. Due to the pandemic, outreach events related to her expedition were cancelled and, instead, translated into digital form. Argueta produced a series of short videos on YouTube titled ``Tiny Ice: Bits from Antarctica.'' The 10-part series, available in English and Spanish, highlights travel, science, and life at the South Pole. Each topic is explained in 2-minute digestible segments, which include photos, videos, and anecdotes to teach viewers about the daily lives of IceCube research team members at the South Pole.

The Tiny Ice videos have been collectively viewed 1,300 times on YouTube. In addition, the videos were shared with the American Family Children's Hospital for their children's enrichment program. At the hospital, which is part of the UW Health network in Wisconsin, inpatients are given a tablet at their hospital bed with a catalog of prerecorded content that helps with the patients' development of growth and play. For some of these patients, it is difficult to participate in live, interactive events due to the nature of their treatments and medications. In order to provide educational and entertaining material for everyone, AFCH sought short videos for all ages and learning styles. We provided them with Argueta's ``Tiny Ice'' videos in English and Spanish, which they were grateful to add to their catalog.

The videos are aligned to Next Generation Science Standards and Polar Literacy Principles. They are intended for use in after-school, formal, and informal learning environments, while also available to the general public. A basic assessment was created to accompany each video in the series.

\section{Summary and Outlook}

The size, scope, and location of the IceCube Neutrino Observatory make it especially challenging to communicate our cutting-edge science to the public. But our strong and continuing relationship with the PolarTREC program and its educators allows the IceCube Collaboration to share the wonders of the IceCube experiment with audiences around the world---even during a pandemic. Our PolarTREC alumni have helped us develop successful education and outreach materials, including curricula for high school students, a virtual tour for an audience of science writers, and a series of bite-sized videos for kids and families in English and Spanish.

In the future, we look forward to continue participating in the PolarTREC program so we can bring more science educators to our unique facility at the South Pole. We intend to continue collaborating with Dr. Shirey, Argueta, and other educators to share the experience of IceCube with audiences of all ages. Over the next few years, as the planet recovers from the COVID-19 pandemic, we are confident that we will be able to work with our PolarTREC partners to adapt to the changing world of science communication, education, and outreach.

\clearpage
\section*{Full Author List: IceCube Collaboration}




\scriptsize
\noindent
R. Abbasi$^{17}$,
M. Ackermann$^{59}$,
J. Adams$^{18}$,
J. A. Aguilar$^{12}$,
M. Ahlers$^{22}$,
M. Ahrens$^{50}$,
C. Alispach$^{28}$,
A. A. Alves Jr.$^{31}$,
N. M. Amin$^{42}$,
R. An$^{14}$
% ...
\\

\noindent
% ...
$^{12}$ Universit{\'e} Libre de Bruxelles, Science Faculty CP230, B-1050 Brussels, Belgium \\
% ...
$^{14}$ Department of Physics and Laboratory for Particle Physics and Cosmology, Harvard University, Cambridge, MA 02138, USA \\
% ...
$^{17}$ Department of Physics, Loyola University Chicago, Chicago, IL 60660, USA \\
$^{18}$ Dept. of Physics and Astronomy, University of Canterbury, Private Bag 4800, Christchurch, New Zealand \\
% ...
$^{22}$ Niels Bohr Institute, University of Copenhagen, DK-2100 Copenhagen, Denmark \\
% ...
$^{28}$ D{\'e}partement de physique nucl{\'e}aire et corpusculaire, Universit{\'e} de Gen{\`e}ve, CH-1211 Gen{\`e}ve, Switzerland \\
% ...
$^{31}$ Karlsruhe Institute of Technology, Institute for Astroparticle Physics, D-76021 Karlsruhe, Germany  \\
% ...
$^{42}$ Bartol Research Institute and Dept. of Physics and Astronomy, University of Delaware, Newark, DE 19716, USA \\
% ...
$^{50}$ Oskar Klein Centre and Dept. of Physics, Stockholm University, SE-10691 Stockholm, Sweden \\
% ...
$^{59}$ DESY, D-15738 Zeuthen, Germany \\
% ...

\subsection*{Acknowledgements}

\noindent
USA {\textendash} U.S. National Science Foundation-Office of Polar Programs,
U.S. National Science Foundation-Physics Division,
U.S. National Science Foundation-EPSCoR,
Wisconsin Alumni Research Foundation,
Center for High Throughput Computing (CHTC) at the University of Wisconsin{\textendash}Madison,
Open Science Grid (OSG),
Extreme Science and Engineering Discovery Environment (XSEDE),
Frontera computing project at the Texas Advanced Computing Center,
U.S. Department of Energy-National Energy Research Scientific Computing Center,
Particle astrophysics research computing center at the University of Maryland,
Institute for Cyber-Enabled Research at Michigan State University,
and Astroparticle physics computational facility at Marquette University;
Belgium {\textendash} Funds for Scientific Research (FRS-FNRS and FWO),
FWO Odysseus and Big Science programmes,
and Belgian Federal Science Policy Office (Belspo);
Germany {\textendash} Bundesministerium f{\"u}r Bildung und Forschung (BMBF),
Deutsche Forschungsgemeinschaft (DFG),
Helmholtz Alliance for Astroparticle Physics (HAP),
Initiative and Networking Fund of the Helmholtz Association,
Deutsches Elektronen Synchrotron (DESY),
and High Performance Computing cluster of the RWTH Aachen;
Sweden {\textendash} Swedish Research Council,
Swedish Polar Research Secretariat,
Swedish National Infrastructure for Computing (SNIC),
and Knut and Alice Wallenberg Foundation;
Australia {\textendash} Australian Research Council;
Canada {\textendash} Natural Sciences and Engineering Research Council of Canada,
Calcul Qu{\'e}bec, Compute Ontario, Canada Foundation for Innovation, WestGrid, and Compute Canada;
Denmark {\textendash} Villum Fonden and Carlsberg Foundation;
New Zealand {\textendash} Marsden Fund;
Japan {\textendash} Japan Society for Promotion of Science (JSPS)
and Institute for Global Prominent Research (IGPR) of Chiba University;
Korea {\textendash} National Research Foundation of Korea (NRF);
Switzerland {\textendash} Swiss National Science Foundation (SNSF);
United Kingdom {\textendash} Department of Physics, University of Oxford.


\begin{thebibliography}{99}

\bibitem{UB2020}
``IceCube’s Upward Bound class goes virtual'' -- article about UWRF's summer 2020 program on IceCube's website: \url{https://icecube.wisc.edu/news/outreach/2020/07/icecube-upward-bound-class-goes-virtual/}.

\bibitem{IAS21}
``IceCube hosts 10-week virtual high school program with Marquette, SDSMT, WIPAC'' -- article about IceCube After School: \\ \url{https://icecube.wisc.edu/news/outreach/2021/04/icecube-hosts-10-week-virtual-high-school-program-with-marquette-sdsmt-wipac/}.

\bibitem{icecubear}
The ICEcuBEAR app is available for free download on Google Play (\url{https://play.google.com/store/apps/details?id=edu.wisc.icecube.icebear}) or at the Apple Store (\url{https://apps.apple.com/us/app/icecubear/id1533578432}).

\bibitem{JA}
``Meet Jocelyn Argueta, our 2019 PolarTREC educator'' -- Q\&A with Jocelyn Argueta on IceCube's website: \url{https://icecube.wisc.edu/news/life-at-the-pole/2019/11/meet-jocelyn-argueta-our-2019-polartrec-educator/}.

\bibitem{JAwebcast}
Recording of IceCube's ``Weekly South Pole Call: Kids' Edition'' with Jocelyn Argueta on April 22, 2020: \url{https://youtu.be/ito1qgK-7nQ}.

\bibitem{JAintro}
Introduction for IceCube's ScienceWriters 2020 presentation by Jocelyn Argueta: \url{https://youtu.be/-HtYiw6cDnU}.

\bibitem{WOtour}
South Pole Tour by John Hardin and Yuya Makino for IceCube Neutrino Observatory at ScienceWriters 2020: \url{https://youtu.be/n9UJOflxNwU}

\bibitem{swya}
Full recording of ``Exploring the Cosmos from the South Pole Ice: A Virtual Tour of the IceCube Neutrino Observatory'' at ScienceWriters 2020: \url{https://youtu.be/oNHzCo9JxDo?t=413}

\end{thebibliography}
\end{document}